\documentclass[aps,amsmath,amssymb,preprint,showpacs,showkeys]{revtex4-1}
\usepackage[b]{esvect}
\usepackage{graphicx}
\usepackage{dcolumn}

\usepackage{bm}
\usepackage{makecell}
\usepackage{footnote}
\usepackage{multirow}
\usepackage{xcolor}
\usepackage[colorlinks=true]{hyperref}
\begin{document}

\title{\boldmath Wave Functions and Leptonic Decays of Bottom Mesons In the
  Relativistic Potential Model}

\author{Hao-Kai Sun}
\email{sunhk@mail.nankai.edu.cn}
\author{Mao-Zhi Yang}
\email{yangmz@nankai.edu.cn}
\affiliation{School of Physics, Nankai University,\\Tianjin, China}

\date{\today}
\begin{abstract}
  We study the wave functions and purely leptonic decays of $b$-flavored mesons
  (pseudoscalars, vector mesons, and higher excited states that are well
  established in experiment) in the relativistic potential model based on our
  previous works. The wave functions are obtained by solving the wave equation
  including the spin-spin and spin-orbit corrections in the effective potential.
  The decay constants of $B$, $B^*$ and some of their excited states that have
  been found in experiment are calculated with these wave functions. Then the
  branching fractions of the purely leptonic decay modes of these bottom mesons
  are studied. Our results are in well agreement with experimental data for decay
  modes that have been measured in experiments. We also provide predictions for
  some yet unmeasured channels, which are useful for experimental test in the
  future.
\end{abstract}

\keywords{leptonic decay}
\pacs{12.39.Pn, 13.20.He, 14.40.Nd}

\maketitle

\section{Introduction}
\label{sec:intro}
Recently, the ATLAS collaboration updated branching ratios of
$\mathcal{B}(B_s\rightarrow\mu^+\mu^-)=(2.8_{-0.7}^{+0.8})\times\! 10^{-9}$ and
$\mathcal{B}(B_d\rightarrow\mu^+\mu^-) < 2.1\times\! 10^{-10}$ with larger
collision data samples~\cite{Aaboud:2018mst} at $95\%$ confidence level. The
former decay channel imposes severe constraints on theoretical study, especially
for physics beyond the Standard Model
(BSM)~\cite{Crivellin:2018gzw,Fleischer:2017ltw}. In general, the purely
leptonic decays with final lepton-neutrino pair or lepton-lepton pair are
considered as rare decays, which have relatively simpler physics than hadronic
decays. Their decay rates are connected straightforwardly with the
Cabibbo-Kobayashi-Maskawa (CKM) matrix elements~\cite{Cabibbo:1963yz,Kobayashi:1973fv}
and the bound-state properties of the bottom meson.

Compared with our previous work~\cite{Sun:2016avp} where only decay constants of
pseudoscalar $B$ and $B_s$ mesons are considered, here we extend our earlier
work by including the decay constants and pure leptonic decays of vector and
higher excited states of bottom mesons. We upgrade the scenario for treating the
energy-momentum conservation for the quark-antiquark inside the bottom mesons.
The theoretical results for the leptonic decays are compared with experimental
data. For the measured decay modes, our prediction are well consistent with
experiment. For the yet unmeasured decay modes, our prediction could be useful
for experimental test in the future.

The paper is organized as followings. In Section~\ref{sec:thf}, we briefly
present the theoretical framework for relativistic potential model, give the
solved wave function for the bottom mesons. The formulas to calculate the decay
constants and branching ratios are also given here. Section~\ref{sec:rlt} is
devoted to numerical results and discussions. Finally, Section~\ref{sec:smy} is
for the conclusion and summary.


\section{Theoretical Framework}
\label{sec:thf}
\subsection{Relativistic potential model and bound-state wave functions}
\label{ssec:rpmwf}
The heavy-light quark-antiquark bound-state systems have been extensively
studied with the relativistic potential model in our previous
works~\cite{Yang:2011ie,Liu:2013maa,Liu:2015lka,Sun:2016avp}. The bound state
wave functions of mesons can be obtained by solving a Schr\"odinger type
equation
\begin{equation}
  (H_0+H')\varPsi(\vv{r})=E\varPsi(\vv{r}),\label{eq:schr}
\end{equation}
where $H_0+H'$ is the effective Hamiltonian, which can be found in
Ref.~\cite{Liu:2015lka} and $E$ is meson's energy. The term $H_0$ reads,
\begin{align}
  H_0=\sqrt{\vv{p}_1^2+m_1^2}+\sqrt{\vv{p}_2^2+m_2^2}+V(r),
  \label{eq:hhvv}
\end{align}
with
\begin{align}
  V(r)=-\tfrac{4}{3}\tfrac{\alpha_S(r)}{r}+br+c,
\end{align}
where $V(r)$ is the effective potential for the strong-interaction between the
quark and anti-quark~\cite{Godfrey:1985xj,Eichten:1978tg,Eichten:1979ms}. The
first term $-\tfrac{4}{3}\tfrac{\alpha_S(r)}{r}$ in $V(r)$ originates from the
one-gluon-exchange diagram for the short distance contributions, and $br$ is for
confinement effects in long distance, while $c$ is a phenomenological parameter
for this heavy-light quark-antiquark system. The other term $H'$ contains
spin-spin hyper-fine interactions and spin-orbit interactions, which are not
given explicitly here (see Ref.~\cite{Liu:2015lka}).

Using the method described in~\cite{Yang:2011ie} and developed
in~\cite{Liu:2013maa,Liu:2015lka}, the wave equation in Eq.~\eqref{eq:schr} can
be solved numerically. The wave functions in momentum space can be written as,
\begin{align}
  &\varPsi_{nlm}(\vv{k})=\varphi_{nl}(k)Y_{lm}(\hat{k}),\label{eq:rwfka}
\end{align}
where the subscripts $nlm$ stands for $n$-th radial wave function ($n=1$ is the
lowest), $l$ orbital angular momentum quantum number ($l=0,1,2,\dots$), and $m$
the magnetic quantum number corresponding to $l$. $\varphi_{nl}(k)$ is the
radial wave functions and $Y_{lm}(\hat{k})$ is the spherical harmonics. The
normalization condition for the wave function is
\begin{align}
  &\int{}dk^3\lvert \varPsi_{nlm}(\vv{k}) \rvert^2=1.\label{eq:rwfkb}
\end{align}

The details for solving the wave equation can be found in our previous works in
Refs.~\cite{Yang:2011ie,Liu:2013maa,Liu:2015lka}, which will not be given here
for briefness. By solving the wave equation numerically, the wave function can
be obtained. In practice, it is convenient to give an analytical form for radial
wave functions by fitting the numerical solution. We find the wave function can
be fitted with the following exponential form,
\begin{equation}
  \varphi_{nl}(\vv{k})=a_1e^{a_2|\vv{k}|^2+a_3|\vv{k}|+a_4}.\label{eq:fitp}
\end{equation}
Next, we give the obtained results for the parameters $a_1\sim a_4$ for each
quantum states.

\begin{figure}[tbp]
  \centering
  \includegraphics[width=0.75\textwidth,origin=l,angle=0]{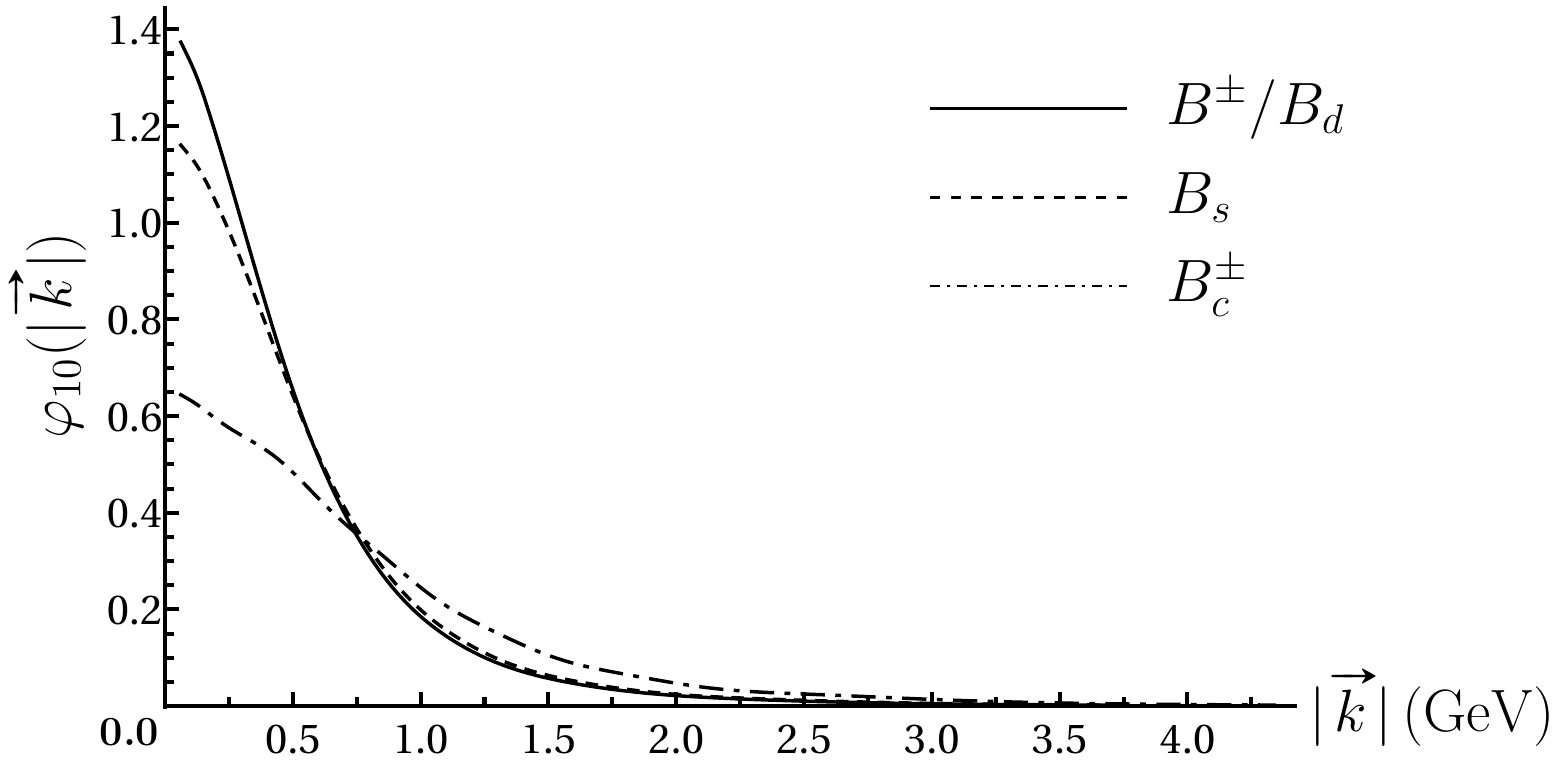}
  \caption{\label{fig:1} Radial wave functions of pseudoscalar mesons,
    normalized by multiplying a constant $Y_{00}=1/\sqrt{4\pi}$.}
\end{figure}

(1) For pseudoscalars $J^P=0^-$, we ignore the difference between the light quark
masses $m_u$ and $m_d$, therefore the radial wave functions of $B^\pm$ and $B_d$
shall be the same. The wave functions of $B^\pm/B_d$, $B_s$, and $B_c^\pm$ are
depicted in FIG.~\ref{fig:1}. It is noted that the differences between
$B^\pm/B_d$ and $B_s$ is relatively smaller than that between $B^\pm/B_d$ and
$B_c^\pm$, since the mass of $c$ quark is greatly larger than that of the
light-quarks. The results for the parameters $a_1,a_2,a_3,a_4$ we obtained are
listed in Table~\ref{tab:aap},
\begin{table}
  \caption{\label{tab:aap}Parameters of fitting functions for pseudoscalar $B$
    mesons.}
  \begin{ruledtabular}
    \begin{tabular}{lrrrr}
      Mesons&$a_1\,(\text{GeV}^{-3/2})$&$a_2\,(\text{GeV}^{-2})$
      &$a_3\,(\text{GeV}^{-1})$&$a_4$\\
      \Xhline{0.5pt}
      $B^\pm/B_d$ &${1.66}_{-0.11}^{+0.07}$&${-1.07}_{-0.16}^{+0.12}$
      &${-0.98}_{-0.12}^{+0.17}$&${-0.13}_{-0.04}^{+0.02}$\\
      $B_s$&${1.97}_{-0.09}^{+0.05}$&${-1.09}_{-0.18}^{+0.12}$
      &${-0.69}_{-0.10}^{+0.07}$&${-0.47}_{-0.07}^{+0.03}$\\
      $B_c$&${1.04}_{-0.03}^{+0.02}$&${-0.51}_{-0.08}^{+0.07}$
      &${-0.44}_{-0.07}^{+0.04}$&${-0.43}_{-0.04}^{+0.02}$
    \end{tabular}
  \end{ruledtabular}
\end{table}

(2) For vector mesons $J^P=1^-$, in our model~\cite{Liu:2013maa,Liu:2015lka}, they
are mixing-states of S-wave ($l=0$) eigenstate $\lvert {}^3S_1 \rangle$ and
D-wave ($l=2$) eigenstate $\lvert {}^3D_1 \rangle$. The S-wave and D-wave wave
functions should be given separately. We use $\varPsi_V^S(\vv{k})$ to denote the
S-wave radial wave function, and $\varPsi_V^D(\vv{k})$ as the D-wave wave
function.

\begin{figure}[tbp]
  \centering
  \includegraphics[width=1.0\textwidth,origin=l,angle=0]{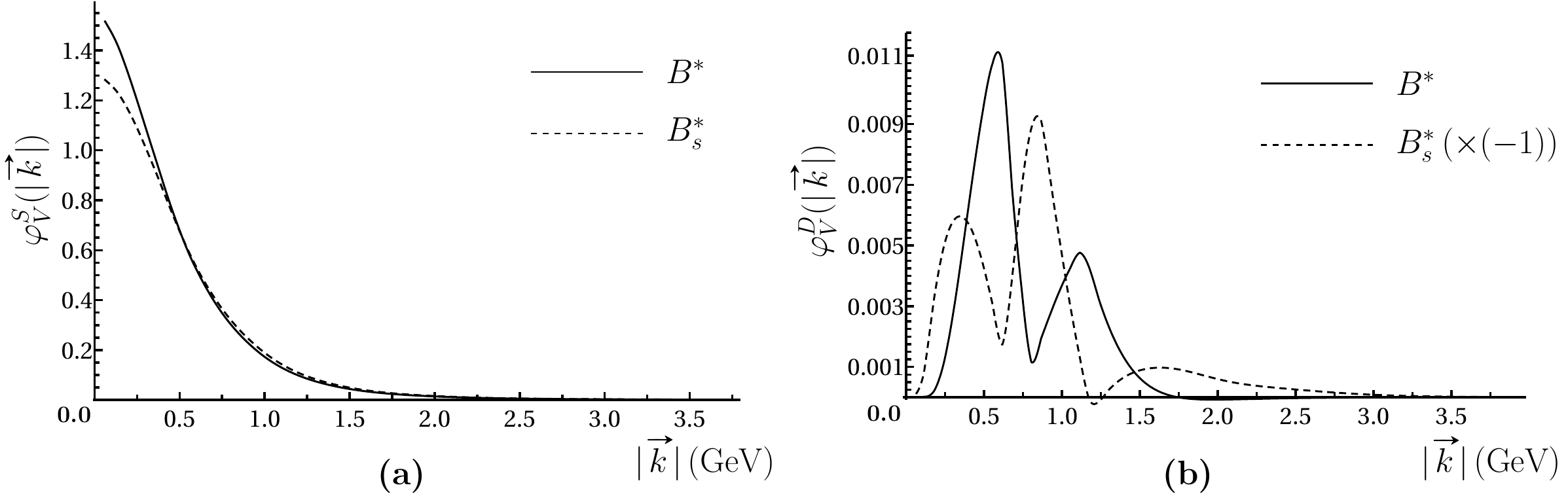}
  \caption{\label{fig:2} Radial wave functions of vector mesons, (a) for the
    S-wave parts and (b) for D-wave parts.}
\end{figure}

The radial wave functions are shown in FIG.~\ref{fig:2}. The contribution of the
D-wave part to the leptonic decay is rather small. Therefore, the fitting
functions are provided only for S-wave part using the same analytic form in
Eq.~\eqref{eq:fitp}, the parameters for vector mesons are collected in
Table~\ref{tab:aav}.

\begin{table}
  \caption{\label{tab:aav}Parameters of fitting functions for vector $B$
    mesons, S-wave part.}
  \begin{ruledtabular}
    \begin{tabular}{lrrrr}
      Mesons&$a_1\,(\text{GeV}^{-3/2})$&$a_2\,(\text{GeV}^{-2})$
      &$a_3\,(\text{GeV}^{-1})$&$a_4$\\
      \Xhline{0.5pt}
      $B^*$ &${1.62}_{-0.08}^{+0.06}$&${-1.09}_{-0.11}^{+0.14}$
      &${-1.23}_{-0.09}^{+0.13}$&${0.03}\!\pm\!{0.01}$\\
      $B_s^*$&${1.45}_{-0.06}^{+0.04}$&${-1.17}_{-0.11}^{+0.15}$
      &${-0.84}_{-0.10}^{+0.09}$&${-0.03}\!\pm\!{0.01}$
    \end{tabular}
  \end{ruledtabular}
\end{table}

\begin{figure}[tbp]
  \centering
  \includegraphics[width=1.0\textwidth,origin=l,angle=0]{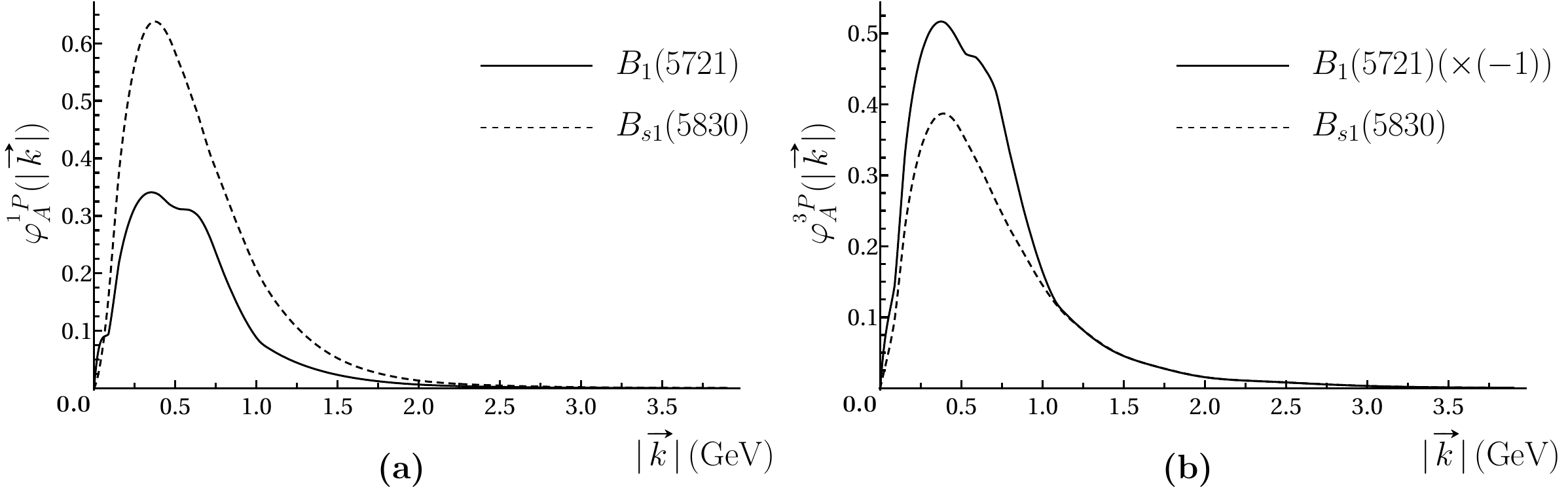}
  \caption{\label{fig:3} Radial wave functions of axial vector ($1^+$) mesons,
    (a) for the ${}^1P_1$ parts and (b) for ${^3}P_1$ parts. }
\end{figure}

\begin{figure}[tbp]
  \centering
  \includegraphics[width=1.0\textwidth,origin=l,angle=0]{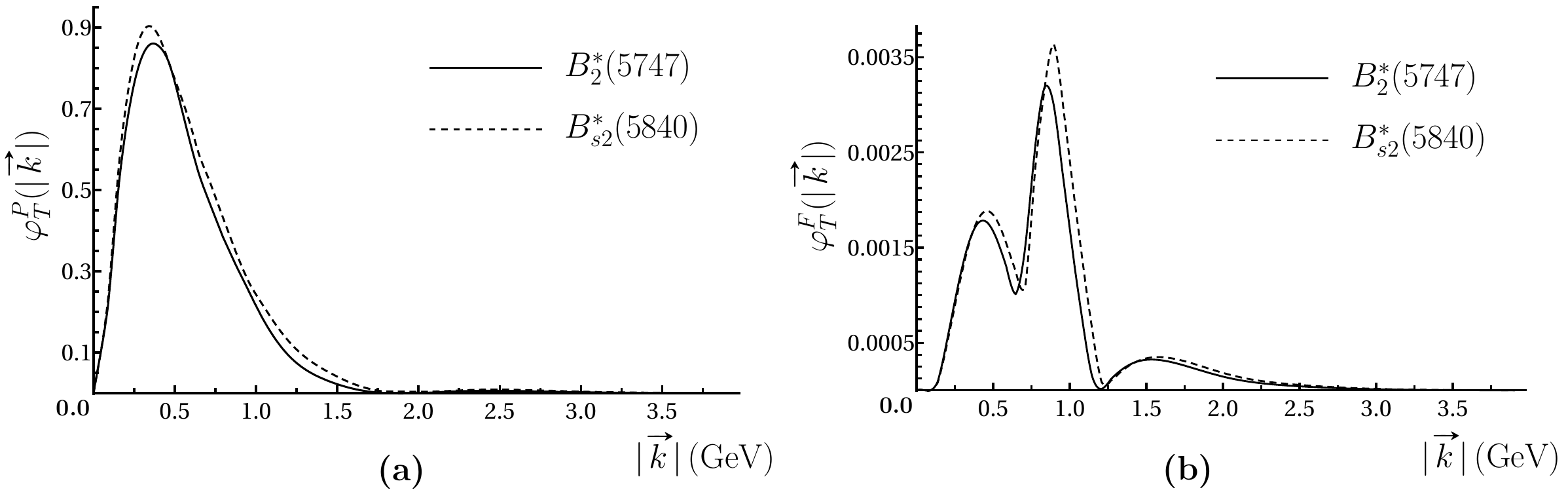}
  \caption{\label{fig:4} Radial wave functions of tensor ($2^+$) mesons, (a) for
    the P-wave parts and (b) for F-wave parts.}
\end{figure}

(3) For higher excited states $1^+$ and $2^+$, only few states
have been found in experiments up to now and information about their decay
channels is very limited. Therefore, we only show the radial wave functions for
$J^P=1^+$ mesons in FIG.~\ref{fig:3} and for $J^P=2^+$ mesons in
FIG.~\ref{fig:4} without giving an analytic form. In principle the numerical
form for the wave functions can be used directly to calculate the meson decays.

\subsection{Decay constants}
\label{ssec:dc}

Using the wave function of the quark-antiquark bound-state, the state of a
bottom meson can be written as~\cite{Leutwyler:1984je,Yang:2011ie}
\begin{equation}
  \lvert M(\vv{P})\rangle=\frac{1}{\sqrt{3N_L}}\sum_i\int{}d^3k_q d^3k_Q
  \delta^{(3)}(\vv{P}-\vv{k}_q-\vv{k}_Q)\varPsi_{nlm}(\vv{k}_q)
  X_{Sm_s}(a_{\vv{k}_Q,s_1}^{i\dagger},b_{\vv{k}_q,s_2}^{i\dagger})
  \lvert 0 \rangle\label{eq:twf}
\end{equation}
where $X_{Sm_s}$ is the spin wave function, $a_{\vv{k}_Q,s_1}^{i\dagger},
b_{\vv{k}_q,s_2}^{i\dagger}$ are creation operators, and $S, m_s, s_1, s_2$ are
the corresponding spin-related quantum numbers. The superscript $i$ is color
index and the normalization factor $N_L$ is obtained via the normalization
condition of the meson state
\begin{align}
  \langle {M(\vv{P})|M(\vv{P}')} \rangle
  &=(2\pi)^32E\delta^{(3)}(\vv{P}-\vv{P}').\label{eq:ncondtb}
\end{align}

The anti-commuting relation of the quark and anti-quark annihilation and
creation operators are
\begin{align}
  \{a_{\vv{k},s}, a_{\vv{k}',s'}^\dagger\}&=\delta_{ss'}\delta^{(3)}(\vv{k}-\vv{k}'),\\
  \{b_{\vv{k},s}, b_{\vv{k}',s'}^\dagger\}&=\delta_{ss'}\delta^{(3)}(\vv{k}-\vv{k}').
  \label{eq:ncondta}
\end{align}

The energy and momentum conservation between the meson and its constituent quark
and antiquark should hold when considering the decays of the bottom mesons. We
take $k_q=(E_q,\vv{k}_q)$, $k_Q=(E_Q,\vv{k}_Q)$, and $P=(m_P,\vv{0})$ as the
four-momenta of the light quark, the heavy quark and the meson in rest frame,
respectively. Due to the energy and momentum conservation, one has
\begin{align}
  E_q + E_Q = m_P,\\
  \vv{k}_q=-\vv{k}_Q=\vv{k}.\label{eq:dcsa}
\end{align}

To keep the four-momentum conservation, the heavy quark is taken off-shell,
while the light quark is kept on-shell in the
Altarelli-Cabibbo-Corbo-Maiani-Martinelli (ACCMM)
scenario~\cite{Altarelli:1982kh,Colangelo:1998eb}. Here we extend the ACCMM
scenario by taking both the light and heavy quark off-shell. The off-shell of
the quarks are a simple treatment for including the energy and momentum carried
by the color field around the quarks. Both the masses of the light and heavy
quarks are taken to be running masses
\begin{align}
  m_q(k)=\sqrt{E_q^2 - \lvert \vv{k}  \rvert^2},\quad
  m_Q(k)=\sqrt{E_Q^2 - \lvert \vv{k}  \rvert^2}.\label{eq:dcsb}
\end{align}
The running masses $m_q(k)$ and $m_Q(k)$ are restricted to be positive in this
work. With Eqs. (\ref{eq:dcsa}) and (\ref{eq:dcsb}), one can obtain
\begin{align}
\sqrt{m_q^2(k)+\lvert \vv{k}  \rvert^2}+\sqrt{m_Q^2(k)+\lvert \vv{k}  \rvert^2}=m_P. \label{eq:ener}
\end{align}
It is not enough to determine the explicit dependence of the running masses
$m_{q,Q}(k)$ on the quark momentum $k$ with the above equation. We assume the
ratio of $m_q(k)/m_Q(k)$ is a fixed parameter in this work, i.e., we define the
ratio for each quark-antiquark pair as
\begin{equation}
  R_i\equiv{}\frac{m_q(\vv{k})}{m_Q(\vv{k})},\quad
  i=(\bar{u}b)\!\sim\!(\bar{d}b),(\bar{s}b),(\bar{c}b).\label{eq:dcsr}
\end{equation}

In our numerical treatment in the following, we find that the fixed ratio $R_i$
can indeed accommodate the experimental data for the measured leptonic decays of
the bottom mesons, and the value of the ratio $R_i$ is approximately around the
ratio of current masses of the light and heavy quarks $m_q/m_b$.

In general, decay constants of pseudoscalar mesons ($B^{\pm},B_d,B_s,B_c^{\pm}$)
are defined as
\begin{equation}
  \langle {0\lvert \bar{q}\gamma_{\mu}\gamma^5 b\rvert} B(P)\rangle
  =if_P P_{\mu}.\label{eq:dcpd}
\end{equation}
Substituting Eq.~\eqref{eq:twf} into the Eq.~\eqref{eq:dcpd}, we
obtain the decay constant
\begin{equation}
  f_P=\sqrt{\frac{3}{(2\pi)^3m_P}}\int{}d^3k\,\varPsi_{100}(\vv{k})
  \left(\sqrt{1+\frac{m_q(k)}{E_q}}\sqrt{1+\frac{m_Q(k)}{E_Q}}-
    \sqrt{1-\frac{m_q(k)}{E_q}}\sqrt{1-\frac{m_Q(k)}{E_Q}}\right).
\label{eq:dcp}
\end{equation}

For vector mesons $B^{*\pm}$, $B^*$, $B_s^*$, there are two types of decay
constants that are defined according to different currents
\begin{equation}
  \langle {0\lvert \bar{q}\gamma_{\mu}b \rvert B^*(P,\epsilon)} \rangle
  =m_{V^*}f_{V^*}\epsilon_{\mu},\quad
  \langle {0\lvert \bar{q}\sigma_{\mu\nu}b \rvert B^*(P,\epsilon)} \rangle
  =-if_{V^*}^T(P_{\mu}\epsilon_{\nu}-\epsilon_{\mu}P_{\nu}),\label{eq:dcvd}
\end{equation}
where $\epsilon_\mu$ is the polarization vector,
$\sigma_{\mu\nu}=\tfrac{i}{2}[\gamma_{\mu},\gamma_{\nu}]$ is the Dirac tensor
matrix. Similarly, the decay constants can be obtained as
\begin{subequations}\label{eq:dcv}
  \begin{align}
    f_{V^*}&=\sqrt{\frac{3}{(2\pi)^3m_{V^*}}}\int{}d^3k\,\varPsi_{n00}'(\vv{k})
    \left(\sqrt{1+\frac{m_q}{E_q}}\sqrt{1+\frac{m_Q}{E_Q}}+\frac{1}{3}
    \sqrt{1-\frac{m_q}{E_q}}\sqrt{1-\frac{m_Q}{E_Q}}\right)
             \label{eq:dcva},\\
     f_{V^*}^T&=\sqrt{\frac{3}{(2\pi)^3m_{V^*}}}\int{}d^3k\,\varPsi_{n00}'(\vv{k})
    \left(\sqrt{1+\frac{m_q}{E_q}}\sqrt{1+\frac{m_Q}{E_Q}}-\frac{1}{3}
    \sqrt{1-\frac{m_q}{E_q}}\sqrt{1-\frac{m_Q}{E_Q}}\right).
             \label{eq:dcvb}
  \end{align}
\end{subequations}
The vector mesons are mixing states of $\lvert n_1 {}^3S_1 \rangle$ and $\lvert
n_2 {}^3D_1 \rangle$ with $n_1,n_2=1,2,3$ (details can be found in
Refs.~\cite{Liu:2013maa,Liu:2015lka}). The wave function
$\varPsi_{n00}'(\vv{k})$ in Eqs. (\ref{eq:dcva}) and (\ref{eq:dcvb}) is the sum
of all the S-wave states. The D-wave states do not contribute to the decay
constants.

\subsection{Branching ratios of purely leptonic decays}
\label{ssec:pld}
We calculate the purely leptonic decays of pseudoscalar and vector $b$-flavored
mesons in this section. The decay modes we consider in this work include
$B_{u(c)}^{\pm}\rightarrow l^{\pm}\nu_l$, $B_{u(c)}^{*\pm}\rightarrow
l^{\pm}\nu_l$, $B_{d(s)}^0\rightarrow l^+l^-$ and $B_{d(s)}^{*0}\rightarrow
l^+l^-$. For the leptonic decays of charged bottom mesons, the decay amplitudes
are dominated by the tree-level diagrams. The branching ratios are calculated to
be
\begin{subequations}\label{eq:pldlv}
  \begin{align}
    \mathcal{B}(B_{q}^{\pm}\rightarrow l^{\pm}\nu_l)=\frac{G_F^2m_l^2M_{B_q}}{8\pi}
    (1-\frac{m_l^2}{M_{B_q}^2})^2f_{B_q}^2\lvert V_{qb}  \rvert^2\tau_{B_q},
    \label{eq:pldlva}\\
    \mathcal{B}(B_{q}^{*\pm}\rightarrow l^{\pm}\nu_l)=\frac{G_F^2M_{B_q^*}^3}{12\pi}
    (1-\frac{3}{2}\frac{m_l^2}{M_{B_q^*}^2}+\frac{1}{2}\frac{m_l^6}{M_{B_q^*}^6})
    f_{B_q^*}^2\lvert V_{qb}  \rvert^2\tau_{B_q^*},
    \label{eq:pldlvb}
  \end{align}
\end{subequations}
where $G_F$ is the Fermi constant, $V_{qb}$ the CKM matrix element,
$M_{B_q^{(*)}}$ and $m_l$ the masses of $B_q^{(*)\pm}$ meson and lepton,
respectively. $\tau_{B_q^{(*)}}$ is the life time of the bottom meson.

For the leptonic decays of the neutral bottom mesons, the decays are induced by
penguin diagrams. The effective Hamiltonian describes such decays
is~\cite{Grinstein:1988me,Grinstein:1990tj,Grinstein:2004vb},
\begin{align}
  \mathcal{H}_{eff}=-\frac{G_F}{\sqrt{2}}\lambda_q\sum_{i=1}^{10}C_i(\mu)Q_i(\mu),\label{eq:heffp}
\end{align}
where $\lambda_q=V_{tb}V_{tq}^*$ and the operators are (we use $b\rightarrow s$
as an example and $b\rightarrow d$ is similar),
\begin{align*}
  &Q_1=(\bar{s}_\alpha c_\beta)_{V-A}(\bar{c}_\beta b_\alpha)_{V-A},
  &Q_2&=(\bar{s}c)_{V-A}(\bar{c}b)_{V-A}\\
  &Q_3=(\bar{s}b)_{V-A}\sum_q (\bar{q}q)_{V-A},
  &Q_4&=(\bar{s}_\alpha b_\beta)_{V-A}\sum_q (\bar{q}_\beta q_\alpha)_{V-A}\\
  &Q_5=(\bar{s}b)_{V-A}\sum_q (\bar{q}q)_{V+A},
  &Q_6&=(\bar{s}_\alpha b_\beta)_{V-A}\sum_q (\bar{q}_\beta q_\alpha)_{V+A}\\
  &Q_7=\frac{\alpha_{em}}{2\pi}m_b\bar{s}_\alpha \sigma^{\mu\nu}(1+\gamma^5)b_\alpha F_{\mu\nu},
  &Q_8&=\frac{\alpha_s}{2\pi}m_b\bar{s}_\alpha \sigma^{\mu\nu}(1+\gamma^5)T_{\alpha\beta}^a
    b_\beta G_{\mu\nu}^a\\
  &Q_9=\frac{\alpha}{2\pi}(\bar{s}b)_{V-A}(\bar{l}l)_V,
  &Q_{10}&=\frac{\alpha}{2\pi}(\bar{s}b)_{V-A}(\bar{l}l)_A\\
\end{align*}
where $\alpha=\tfrac{e^2}{4\pi}$ is the electromagnetic coupling constant.
Except for the contribution of the operators $Q_7$ and $Q_9$, the operators
$Q_{1-6,8}$ also contribute to the decay process $B_{d(s)}^{*0}\rightarrow
l^+l^-$ up to next-to-leading (NLO) order in $\alpha_s$ expension in QCD. The
contributions from the operators $Q_{1-6,8}$ can be absorbed by a redefinition
of two effective Wilson coefficients $C_{7,9}\to C_{7,9}^{eff}$. The explicit
form of $C_{7,9}^{eff}$ can be found in
Refs.~\cite{Grinstein:2004vb,Seidel:2004jh,Greub:2008cy}, we do not repeat it
here.

Next we give the branching ratio of the pure leptonic decay of the neutral
bottom mesons. The branching ratio of $B_{d(s)}^0\rightarrow l^+l^-$ is
~\cite{Bobeth:2013uxa}:
\begin{equation}
  \bar{\mathcal{B}}(B_q\rightarrow l^+l^-)=
  \frac{G_F^2M_{B_q}m_l^2\alpha^2}{4\pi^3\Gamma_H^q}f_{B_q}^2\lvert \lambda_q  \rvert^2
  \sqrt{1-\frac{4m_l^2}{M_{B_q}^2}}\,\lvert C_{10} \rvert^2.\label{eq:pldllp}
\end{equation}
The hat over $\mathcal{B}$ indicates that it is the averaged time-integrated
branching ratio that depends on the details of $B_q^0-\bar{B_q^0}$
mixing~\cite{DeBruyn:2012wk}. $\Gamma_H^q$ denotes the total decay width of the
heavier mass-eigenstate . In Ref.~\cite{Bobeth:2013uxa}, the authors define a
different Wilson coefficient $C_A$ and its relation to $C_{10}$ in
Eq.~\eqref{eq:heffp} is straightforward: $\lvert C_A \rvert = \tfrac{\sin^2
  \theta_W}{2}\lvert C_{10} \rvert$, where $\theta_W$ is the weak-mixing angle
(the Weinberg angle).

For the process of the vector meson decaying into charged lepton-anti-lepton pair,
the branching ratio reads~\cite{Grinstein:2015aua},
\begin{equation}
  \bar{\mathcal{B}}(B_q^*\rightarrow l^+l^-)=\frac{1}{\Gamma_{B_q^*}}
  \frac{G_F^2M_{B_q^*}^3\alpha^2}{96\pi^3}f_{B_q^*}^2\lvert \lambda_q  \rvert^2
  \left(\Bigl\lvert C_9^{eff} + 2\frac{m_bf_{B_q^*}^T}{M_{B_q^*}f_{B_q^*}}C_7^{eff} \Bigr\rvert^2
    +\lvert C_{10} \rvert^2\right),\label{eq:pldllv}
\end{equation}
where the contributions of order $\mathcal{O}(m_l^2/M_{B_q^*}^2)$ are neglected
and $m_c \ll m_b$ is considered.

\section{Numerical Calculation}
\label{sec:rlt}
In this section, we calculate the decay constants in Eqs. \eqref{eq:dcp},
\eqref{eq:dcv}, and leptonic decay branching ratios in Eqs. \eqref{eq:pldlv}, \eqref{eq:pldllp},
\eqref{eq:pldllv} numerically, and compare them with experimental measurements.

The parameters used in this work are collected in Table~\ref{tab:i}, which are
quoted from PDG~\cite{Tanabashi:2018oca}.

\begin{table}[tbp]
  \caption{\label{tab:i} Numerical inputs.}
  \begin{ruledtabular}
  \begin{tabular}{llc|llc}
    Parameter&Value&Unit\,\,
    &Parameter&Value&Unit\\
    \Xhline{1.0pt}
    $G_F$&$1.16638\times 10^{-5}$&$\text{GeV}^{-2}$
    &$\lvert V_{ub} \rvert$&$3.94(36)\times 10^{-3}$&--\\
    $\alpha_s^{(5)}(M_z)$&$0.1181(11)$&--
    &$\lvert V_{cb} \rvert$&$4.22(8)\times 10^{-2}$&--\\
    $\alpha^{(5)}(M_Z)$&$1/127.955(10)$&--
    &$\lvert V_{td} \rvert$&$8.1(5)\times 10^{-3}$&--\\
    $M_Z$&$91.1876(21)$&GeV
    &$\lvert V_{ts} \rvert$&$3.94(23)\times 10^{-2}$&--\\
    $M_t$&$173.1(9)$&GeV
    &$\lvert V_{tb} \rvert$&$1.019(25)$&--\\
    \hline
    $M_{B^\pm}$&$5279.32(14)$&MeV
    &$M_{B^*}$&$5324.65(25)$&MeV\\
    $M_{B_s}$&$5366.89(19)$&MeV
    &$M_{B_s^*}$&$5415.4_{-1.5}^{+1.8}$&MeV\\
    $M_{B_d}$&$5279.63(15)$&MeV
    &$M_{B_c}$&$6274.9(8)$&MeV\\
    $\tau_{B^\pm}$&$1.638(4)$&ps
    &$\tau_{B_d}$&$1.520(4)$&ps\\
    $2/(\Gamma_{B_{sL}}\!\!+\!\Gamma_{B_{sH}})$&$1.509(4)$&ps
    &$\Gamma_{B_{sL}}\!\!-\!\Gamma_{B_{sH}}$&$0.088(6)$&$\text{ps}^{-1}$\\
  \end{tabular}
  \end{ruledtabular}
\end{table}

(1) For pseudoscalar $B$  mesons ($J^P=0^-$)

The $J^P=0^-$ pseudoscalar meson is $\lvert 1{}^1S_0 \rangle$ state. Besides the
parameters in Table~\ref{tab:i}, additional numerical input is the Wilson
coefficient $C_{10}$ whose expression is known up to next-to-leading-order (NLO)
in electro-weak (EW) corrections~\cite{Bobeth:2013tba} and
next-next-to-leading-order (NNLO) in QCD~\cite{Hermann:2013kca}. Its value is
$\lvert C_{10}(\mu_b=m_b\sim 5 \text{GeV}) \rvert=4.053\pm 0.032$.

\begin{table}
  \caption{\label{tab:ii}Decay constants, purely leptonic branching ratios of
    pseudoscalars.}
  \begin{ruledtabular}
  \begin{tabular}{ccclrr}
    {\bfseries B-Meson\,}&$R_i$~(Eq.\eqref{eq:dcsr})\,&$f_B\,(\text{MeV})$\,
    &{\bfseries channel}&{\bfseries this work}&{\bfseries Exp.}\\
    \Xhline{1.0pt}
    \multirow{6}{*}[0ex]{$(bq)$}
      &\multirow{6}{*}[0ex]{$1.0\times 10^{-3}$} &\multirow{6}{*}[0ex]{$210(10)$}
          &$B^+\rightarrow{}e^+\nu_e$&$1.27(26)\times\! 10^{-11}$
          &$<9.8\times\! 10^{-7}$~\cite{Satoyama:2006xn}\\
      & &
          &$B^+\rightarrow{}\mu^+ \nu_{\mu}$ &$5.4(1.1)\times\! 10^{-7}$
          &$<10.7\times\! 10^{-7}$~\cite{Sibidanov:2017vph}\\
      & &
          &$B^+\rightarrow\tau^+\nu_{\tau}$&\quad{}$1.21(25)\times\! 10^{-4}$
          &\quad{}$1.25(28)(27)\times\! 10^{-4}$~\cite{Kronenbitter:2015kls}\\
    \Xcline{4-6}{0.6pt}
      & &
          &$B_d\rightarrow{}e^+e^-$&$3.00(49)\times\! 10^{-15}$
          &$<8.3\times\! 10^{-8}$~\cite{Aaltonen:2009vr}\\
      & &
          &$B_d\rightarrow{}\mu^+\mu^-$&$1.28(21)\times\! 10^{-10}$
          &$<2.1\times\! 10^{-10}$~\cite{Aaboud:2018mst}\\
      & &
          &$B_d\rightarrow{}\tau^+\tau^-$&$2.68(44)\times\! 10^{-8}$
          &$<1.6\times\! 10^{-3}$~\cite{Aaij:2017xqt}\\
    \Xhline{1.0pt}
    \multirow{3}{*}[0ex]{$(bs)$}
      &\multirow{3}{*}[0ex]{$1.0\times 10^{-2}$}&\multirow{3}{*}[0ex]{$229(11)$}
          &$B_s\rightarrow{}e^+e^-$&$8.1(1.3)\times\! 10^{-14}$
          &$<2.8\times\! 10^{-7}$~\cite{Aaltonen:2009vr}\\
      & &
          &$B_s\rightarrow{}\mu^+\mu^-$&$3.45(55)\times\! 10^{-9}$
          &$2.8_{-0.7}^{+0.8}\times\! 10^{-9}$~\cite{Aaboud:2018mst}\\
      & &
          &$B_s\rightarrow{}\tau^+\tau^-$&$7.3(1.2)\times\! 10^{-7}$
          &$<5.2\times\! 10^{-3}$~\cite{Aaij:2017xqt}\\
    \Xhline{1.0pt}
    \multirow{3}{*}[0ex]{$(bc)$}
      &\multirow{3}{*}[0ex]{$0.3$}&\multirow{3}{*}[0ex]{$429(21)$}
          &$B_c^+\rightarrow{}e^+\nu_e$&$2.24(24)\times\! 10^{-9}$
          &--\\
      & &
          &$B_c^+\rightarrow{}\mu^+ \nu_{\mu}$&$9.6(1.0)\times\! 10^{-5}$
          &--\\
      & &
          &$B_c^+\rightarrow{}\tau^+ \nu_{\tau}$&$2.29(24)\times\! 10^{-2}$
          &--\\
  \end{tabular}
  \end{ruledtabular}
\end{table}

Our numerical results are listed in Table~\ref{tab:ii}. The uncertainties of
branching ratios mainly come from decay constants (whose relative error is $5\%$
in our scenario) and CKM matrix elements (see the full review in
paper~\cite{Tanabashi:2018oca}). For the two well-measured decay modes
$B^+\rightarrow\tau^+\nu_\tau$ and $B_s\rightarrow \mu^+\mu^-$, our results are
in good agreement with experiment data. The other predicted branching ratios are
all consistent with experimental data. They are well below the experimental
upper limit. The branching ratios of $B^+\rightarrow{}\mu^+ \nu_{\mu}$ and
$B_d\rightarrow{}\mu^+ \mu^-$ are very near to the present experimental upper
limit, which are very hopeful to be detected with the upgraded detectors at
Belle II and/or LHCb ~\cite{Albrecht:2017odf} in the near future.

In addition, the prediction of the branching ratio of
$B_c^+\rightarrow\tau^+\nu_\tau$ is two orders of magnitude larger than that of
$B^+$. Therefore it could be a possible channel to be measured in experiments in
future.

(2) For vector $B$ mesons $J^P=1^-$

Since the total decay widths of vector $B$-meson are not well measured in
experiments up to now, in this work, their values of theoretical estimations will
be used. As stated in Refs.~\cite{Chang:2018sud,Choi:2007se,Cheung:2014cka},
vector $B$-mesons' decays are dominated by the electromagnetic processes
$B^*\rightarrow B\gamma$ and thus we make the assumption that the total decay
width $\Gamma$ approximately equals $\Gamma(B^*\rightarrow B\gamma)$. The
respective values are $\Gamma_{B_s^*}\sim{}0.068(18)\,\text{KeV}$,
$\Gamma_{B_d^{*}}\sim{}0.148(20)\,\text{KeV}$, and
$\Gamma_{B^{*+}}\sim{}0.468_{-0.075}^{+0.073}\,\text{KeV}$. The ``running'' mass
of b-quark is $m_b(\bar{MS})=4.18\,\text{GeV}$. The values of effective Wilson
coefficients are
$C_9^{eff}=C_9^{eff}(m_b,m_{B^*}^2)\sim{}C_9^{eff}(m_b,m_{B_s^*}^2)=4.560+i0.612$
and
$C_7^{eff}=C_7^{eff}(m_b,m_{B^*}^2)\sim{}C_7^{eff}(m_b,m_{B_s^*}^2)=-0.384-i0.111$
\cite{Grinstein:2015aua}.

The results for the branching ratios of vector B-meson are shown in
Table~\ref{tab:iii}. The uncertainties mainly come from the uncertainties of the
decay width of vector mesons and CKM parameters. In general, the branching
ratios of the leptonic decay of vector B-mesons are very small.

\begin{table}
  \caption{\label{tab:iii}Decay constants, purely leptonic branching ratios of
    vectors with $l=e,\mu$.}
  \begin{ruledtabular}
  \begin{tabular}{cccclr}
    {\bfseries B-Meson}\,\,&$R_i$~(Eq.\eqref{eq:dcsr})\,&$f_{B_q^*}\,(\text{MeV})$\,
    &$f_{B_q^*}^T\,(\text{MeV})$&{\bfseries channel}&{\bfseries this work}\\
    \Xhline{1.0pt}
    \multirow{4}{*}[0ex]{$(bq)$}
      &\multirow{4}{*}[0ex]{$1.0\times 10^{-3}$} &\multirow{4}{*}[0ex]{$223(16)$}
      &\multirow{4}{*}[0ex]{$201(14)$}
          &$B^{*+}\rightarrow{}e^+\nu_e$ &$9.0(2.5)\times\! 10^{-10}$\\
      & & & &$B^{*+}\rightarrow{}\mu^+ \nu_{\mu}$ &$9.0(2.5)\times\! 10^{-10}$\\
      & & & &$B^{*+}\rightarrow\tau^+\nu_{\tau}$ &$7.5(2.1)\times\! 10^{-10}$\\
    \Xcline{5-6}{0.6pt}
      & & & &$B_d^*\rightarrow{}l^+l^-$&$3.16(77)\times\! 10^{-13}$\\
    \Xhline{1.0pt}
    $(bs)$&$1.0\times 10^{-2}$&$242(17)$&$219(15)$
        &$B_s^*\rightarrow{}l^+l^-$ &$2.02(64)\times\! 10^{-11}$\\
  \end{tabular}
  \end{ruledtabular}
\end{table}


\section{Summary}
\label{sec:smy}
To summarize, we study the leptonic decays of b-flavored mesons in the
relativistic potential model. The decay constants of the bottom mesons and
branching ratios of the leptonic decay modes are calculated. The predictions for
the branching ratios of $B^+\rightarrow\tau^+\nu_\tau$ and $B_s\rightarrow
\mu^+\mu^-$ are well consistent with the experimental measurements. The other
predicted branching ratios are well below the experimental upper limit. For
$B^+\rightarrow{}\mu^+ \nu_{\mu}$ and $B_d\rightarrow{}\mu^+ \mu^-$ decays,
the predictions of the branching ratios are very near to the present experimental
upper limit, which are very hopeful to be detected with the upgraded detectors
at Belle II and/or LHCb in the near future. For the leptonic decays of the
vector B-mesons, the branching ratios are very small.


\acknowledgments

This work is supported in part by the National Natural Science
Foundation of China under Contract Nos. 11875168 and 11375088.

%
\end{document}